\documentclass[
aps,%
12pt,%
final,%
notitlepage,%
oneside,%
onecolumn,%
nobibnotes,%
nofootinbib,%
superscriptaddress,%
noshowpacs,%
centertags]%
{revtex4}

\usepackage{epsf}

\usepackage{amsmath}
\usepackage{graphicx}
\usepackage[cp866]{inputenc}
\usepackage[english]{babel}

\newcommand{\be}{\begin{equation}}
\newcommand{\ee}{\end{equation}}
\newcommand{\bea}{\begin{equation}\begin{aligned}}
\newcommand{\eea}{\end{aligned}\end{equation}}

\newcommand{\beq}{\begin{eqnarray}}
 \newcommand{\eeq}{\end{eqnarray}}

\def\fun#1#2{\lower3.6pt\vbox{\baselineskip0pt\lineskip.9pt
\ialign{$\mathsurround=0pt#1\hfil ##\hfil$\crcr#2\crcr\sim\crcr}}}

\newcommand{{\SD}}{\rm SD}

\begin{document}


\title{Hydrogen recombination in the early Universe in the presence of a magnetic field}
\author{N.O. Agasian}
\affiliation{
National Research Nuclear University MEPhI, \\
Moscow, Russia}
\affiliation{
Institute of Theoretical and Experimental Physics,\\
Moscow, Russia}
\author{S.M. Fedorov}
\affiliation{
Institute of Theoretical and Experimental Physics,\\
Moscow, Russia}

\begin{abstract}
Hydrogen recombination in the early Universe in the presence of a magnetic field is studied.
An equation for the temperature of recombination in the presence of a magnetic field is derived.
Limiting cases of weak and strong fields are considered.
It is demonstrated that there exists a critical magnetic field,
above which the system stays in the phase of atomic hydrogen for all temperatures.
The relative shift of the temperature of recombination in the presence
of a magnetic field is estimated and it is demonstrated that this shift is small.
\end{abstract}

\maketitle

It is well known that in the early Universe a phase transition from the plasma of electrons
and protons to the atomic hydrogen took place at the recombination temperature $T_r=0.27$~eV.
At the same time there exist magnetic fields in the universe,
and magnetic fields were also present at the recombination time.
Thus it is of interest to study the influence of the magnetic fields on the process
of hydrogen recombination from electron-proton plasma.
As shown below, in the "weak" magnetic field it is sufficient
to consider the effect of the field on electronic component of the plasma only.
The contribution of protons and hydrogen atoms to the recombination temperature
is small compared to the contribution of electrons.
Since the recombination temperature is much less then the electron mass, $T_r\ll m_e$,
we will further consider the non-relativistic approximation.

Electron density in the magnetic field in non-relativistic approximation is given by
\be
n_e = e ^{\frac{\mu_e}{T}}
\left( \frac{eB}{2\pi} \right) \sum^\infty_{n=0} \sum_{s=\pm 1}
\int^{+\infty}_{-\infty} \frac{ d p_z}{2\pi} e^{{-\omega_B} /T},
\label{1}
\ee
where $\mu_e$ is the chemical potential and the spectrum of electrons in a magnetic field is
\be
\omega^2_B = p^2_z +
m^2_e +eB (2n+1+s) \label{2} \ee
Next we find that ($z\equiv e^{{\mu_e}/{T}}$)
\be n_e = z\frac{eBT}{2\pi^2} \sum^\infty_{n=0}
\sum_{s=\pm1 }\int^\infty_0 dx e^{-\sqrt{x^2+ \lambda^2_T}},
\label{3}
\ee
here we use the following notation
\be
\lambda^2_T = \frac{
m^2_e + eB (2n+1+s)}{T^2} \label{4} \ee
Since
 \be
 \int^\infty_{\lambda_T} \frac{ xdx}{\sqrt{x^2-\lambda^2_T}} e^{-x} =
 \lambda_T K_1 (\lambda_T),
 \label{5}
 \ee
where $K_1$ is the Macdonald fuction, we find that
 \be
 n_e = z \frac{eBT}{2\pi^2} \sum^\infty_{n=0}
\sum_{s=\pm 1} \lambda_T K_1 (\lambda_T). \label{6} \ee
Summing up the spin states $s=\pm1$ and keeping the contribution $n=0$ separately we obtain
\be
n_e = z \frac{eB}{2\pi^2} m_e K_1 \left( \frac{m_e}{T}\right)
 + 2 z \frac{eB}{2\pi^2} \sum^\infty_{n=1} \sqrt{m^2_e + 2 n eB} K_1 \left( \frac{ \sqrt{ m^2_e +2
neB}}{T}\right)
\label{7}
\ee

Let us consider the limit of weak field, $eB\ll T^2$. In this limit one can use the Euler-Maclaurin summation formula
\be
\sum^\infty_{n=1} \Phi (n) = -\frac12 \Phi (0) + \int^\infty_0 dz
\Phi(z) - \frac{1}{12} \Phi' (0)
\label{8}
\ee
Then the electron density in the weak field takes the form
\be
n_e = z \frac{ m^2_e T}{\pi^2} K_2 \left( \frac{m_e}{T} \right)
 - z
\frac{(eB)^2}{24\pi^2 m_e} \left[ 2K_1 \left( \frac{m_e}{T}
\right) - \frac{m_e}{T} K_0 \left( \frac{ m_e}{T} \right) -
\frac{m_e}{T} K_2 \left( \frac{m_e}{T}\right) \right]
\label{9}
\ee
Taking into account the properties of Macdonald function
\be -xK_0 (x) + xK_2 (x)
= 2 K_1 (x)
\label{10}
\ee we find
\be n_e =z \frac{ m^2_e
T}{\pi^2} K_2 \left( \frac{m_e}{T} \right) + z
\frac{(eB)^2}{12\pi^2 T} K_0 \left( \frac{m_e}{T} \right)
\label{11}
\ee
Taking into account that $K_\nu (x)\rightarrow
\sqrt{{\pi}/{2x}} e^{-x}$ when $x\gg 1$ we obtain the following expression
for the electron density in the non-relativistic limit in the weak magnetic field
\be n_e = 2z \left( \frac{ m_e T}{2\pi}\right)^{3/2}
e^{-m_e/T} \left[ 1+ \frac{(eB)^2}{12 m^2_e T^2} \right]
\label{13}
\ee

To derive Saha equation in the presence of magnetic field we will single
out the value of electron density in the absence of magnetic field
$$
n_e = n_0 \varphi (B)
$$
\be
\varphi (B)= \frac{eB }{2m_eT}
 +eB\sum^\infty_{n=1} \frac{ (m^2_e + 2n eB)^{1/4}}{T
m_e^{3/2}} \exp \left\{ \frac{ m_e - \sqrt{ m^2_e + 2
neB}}{T}\right\} \label{14} \ee
where
\be
 n_0 = 2 e ^{\frac{\mu_e}{T}} \left( \frac{m_eT}{2\pi} \right)^{3/2}
 e^{-m_e/T}
 \label{16}
 \ee
is the value of electron density $n_e$ at $B=0$.
Densities of protons and hydrogen atoms in non-relativistic approximation are
\be
 n_p = 2 \left( \frac{ m_pT}{2\pi} \right)^{3/2}
 e^{(\mu_p-m_p)/T}
 \label{17}
 \ee

  \be
  n_H = 4 \left( \frac{ m_HT}{2\pi}\right)^{3/2}
  e^{(\mu_H-m_H)/T}
  \label{18}
  \ee
Here $\mu_p$ and $\mu_H$ are chemical potentials of protons and hydrogen atoms. We will further use the notations
  \be
  X_p \equiv \frac{ n_p}{n_B}, ~~ X_H \equiv\frac{n_H}{n_B},
  \ee
where $n_B$ is the baryon density. Then the baryon number density
$n_B(T) = \eta_B n_\gamma (T)$ and the density of the number of photons is the known function of temperature
\be n_\gamma = 2 \frac{
\zeta(3)}{\pi^2} T^3, ~~ \zeta (3) \simeq 1.2 \label{19} \ee
The value of baryon-photon ratio is $\eta_B \simeq 6.65 \cdot
10^{-10}$.

Following the derivation of Saha equation in the absence of the magnetic field, presented in~\cite{Rubakov},
we will obtain the corresponding equation with the account for the magnetic field
\be X_p+ n_B X^2_p \left(
\frac{2\pi}{m_eT}\right)^{3/2} \frac{1}{\varphi (B)}
e^\frac{\Delta_H}{T} =1,
\label{20}
\ee
where $\Delta_H \equiv m_p +
m_e-m_H =13.6$ eV is the biding energy of hydrogen and we have neglected the difference between $m_p$ and $m_H$.
Expressing $n_B$ through baryon-photon ratio $\eta_B$ and using the formula for $n_\gamma$,
we obtain the equation with dimensionless variables only
\be
X_p
+ \frac{2\zeta(3)}{\pi^2} \eta_B \left( \frac{2\pi
T}{m_e}\right)^{3/2} X_p^2 \frac{1}{\varphi(B)} e^{\frac{\Delta_H}{T}}
=1
\label{21}
\ee
The second term is a relative concentration of hydrogen atoms, expressed through $X_p$,
\be
X_H=\frac{2\zeta(3)}{\pi^2} \eta_B \left( \frac{2\pi
T}{m_e}\right)^{3/2} X_p^2 \frac{1}{\varphi(B)} e^{\frac{\Delta_H}{T}}
\label{21a}
\ee
Recombination takes place when both $X_p\sim 1$, $X_H\sim 1$.
This leads to
\be
\frac{\Delta_H}{T_B} =-{\rm ln} \left( \frac{2\zeta(3)}{\pi^2}
\eta_B \left( \frac{2\pi T_B}{m_e}\right)^{3/2} \frac{1}{\varphi}
\right)
\label{22}
\ee
This formula is obtained by substituting $X_p(T_B)\sim X_H(T_B)\sim 1$ into the formula~(\ref{21a}), and it allows to
find the recombination temperature in a magnetic field $T_B$. Let us rewrite this equation in the following form
\be \frac{
\Delta_H}{T_B} = {\rm ln} \left( \frac{\pi^2}{2\zeta(3)}
\eta^{-1}_B \left( \frac{m_e}{2\pi}\right)^{3/2} \left(
\frac{\Delta_H}{T_B}\right)^{3/2}
\frac{1}{\Delta_H^{3/2}}\varphi(B)\right)
\label{23}
\ee Equation can be expressed as
\be x={\rm ln} A x^\alpha \varphi(B), \label{24}
\ee where \be A= \frac{\sqrt{\pi}}{4\sqrt{2}\zeta(3)}
\left(\frac{m_e}{\Delta_H} \right)^{3/2} \eta^{-1}_B, ~~
\alpha=\frac32,~~x=\frac{\Delta_H}{T_B}
\label{25}
\ee In the absence of a magnetic field,
$\varphi(B)=1$, the solution is given by~\cite{Rubakov}
\be T_r =\frac{
\Delta_H}{{\rm ln} \left( \frac{\sqrt{\pi}}{4\sqrt{2}\zeta(3)}
\left(\frac{m_e}{\Delta_H} \right)^{3/2} \eta^{-1}_B\right)}
\label{26}
\ee
From~(\ref{26}) we find $T_r\simeq 0.38$ eV.
The numerical solution of the equation~(\ref{24}) with $\varphi(B)=1$ gives
$T_r=0.33$ eV~\cite{Rubakov}.

Let us consider the temperature of recombination in the case of weak field $eB\ll T^2$.
From equation~(\ref{13}) one has for $\varphi(B)$
\be
\varphi=1+ \frac{(eB)^2}{12 m^2_e T^2}
\label{27}
\ee

To find the correction to the temperature of recombination in this limit, one has to solve
the equation~(\ref{24}) with $\varphi(B)$ defined by the expression~(\ref{27}) using
the perturbation theory, $T_B = T_r (1-\varepsilon)$ with $\varepsilon\ll 1$.
We find
\be T_B = T_r
\left(1-\frac{(eB)^2 T_r^3}{12 m^2_e \Delta^5_H}\right), ~~
\sqrt{eB}\ll T
\label{28}
\ee

Let us consider another limiting case of "strong" magnetic field
$m_e^2\gg eB\gg m_eT$. In this limit we can leave only the lowest Landau level in $\varphi(B)$.
The electron density is then given by
\be n_e = n_0
\frac{eB}{2m_eT} = n_0 \frac{eB}{2m_e\Delta_H} x, ~~
x=\frac{\Delta_H}{T}
\label{29}
\ee
From equation~(\ref{24}) we find
\be x={\rm ln} A \frac{eB}{2m_e\Delta_H} x^{\alpha+1}
\label{30}
\ee
The solution of this equation is given by
\be T_B = \frac{T_r}{1+
\frac{T_r}{\Delta_H} {\rm ln} \frac{eB}{2m_e\Delta_H}} , ~~ m^2_e
\gg eB \gg m_eT
\label{31}
\ee
The numerical solution of equation ~(\ref{24}) is shown in Fig.~(\ref{fig_lambda_t}).
Dashed line corresponds to unphysical solution that goes from zero. As one can see,
there exists a critical value of a magnetic field, $eB_c/m_e\Delta_H=0.0462$
($eB_c=1.67\cdot10^7$ Gs), above which the solution does not exit. This means that the system is
in the state of atomic hydrogen at $B>B_c$.
In such magnetic fields electrons occupy the lowest Landau level, and we always
have the bound system of proton and electron. Note that the critical magnetic field is
$eB_c\ll eB=m_e^2\sim 10^{13}$ Gs and the constraints on the magnetic field used above are valid.

\unitlength=1pt
\begin{figure}[!ht]
\begin{picture}(220,200)
\put(0, 20){\includegraphics{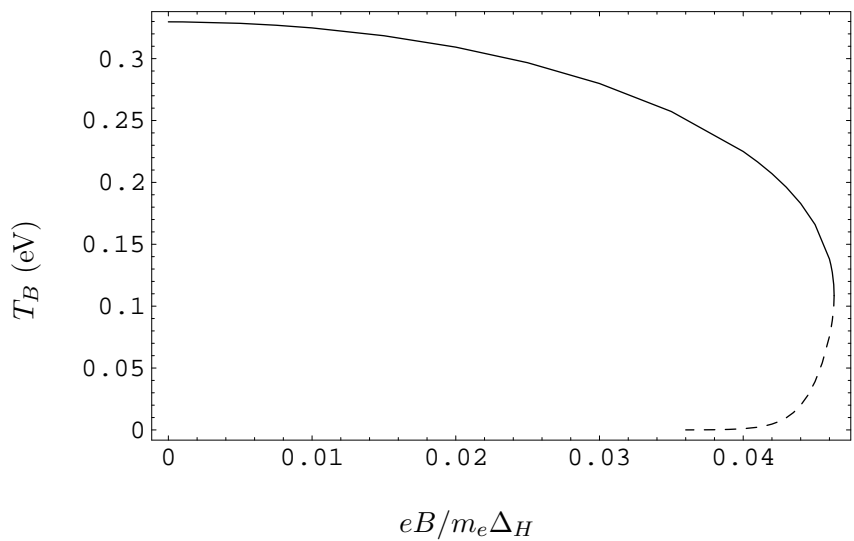}}
\put(120, 0){$eB/{m_e\Delta_H}$}
\put(-25, 80){\rotatebox{90}{$T_B$ (eV)}}
\end{picture}
\caption{Recombination temperature as a function of the magnetic field. Dashed line corresponds to an unphysical branch of the solution.}
\label{fig_lambda_t}
\end{figure}

Let us consider the corrections due to the change of densities of protons and hydrogen atoms in a magnetic field.
Clearly, the correction to the recombination temperature due to protons in a weak field is given by~(\ref{28})
with the substitution $m_e \to m_P$.
One can see that this correction is strongly suppressed as compared to the correction due to electrons, the suppression
is due to proton mass.

The hydrogen density in the weak fields $eB\ll T^2$ is in the first order due to the shift of
binding energy $\Delta_H$. In the weak fields this shift is well known~\cite{Landau}, $\Delta
E=B^2/12m_e^2e^2$.
Thus, the relative change of hydrogen density compared to the change of electron density is
$(\Delta n_H/n_H)/(\Delta n_e/n_e)=T_r/m_ee^4\approx 0.01$.

Let us make an estimate of a magnetic field at the recombination time. Homogeneous magnetic field
at time $t_i$ depends on the scale factor $a_i$~\cite{Grasso:2000wj}
\be
B(t)=B(t_i)\left(\frac{a(t_i)}{a(t)}\right)^2
\label{32}
\ee

On the other hand, at the stage of dominating nonrelativistic matter, the effective
temperature is $T_{\rm eff}\propto 1/a^2(t)$~\cite{Rubakov}.
Thus we have an estimate for the magnetic field at the recombination stage
\be
B_r=\left(\frac{T_r}{T_0}\right)B_0,
\label{3}
\ee
where $B_0$ and $T_0$ are modern values of magnetic field and temperature.
Setting $T_r=0.27$ eV, $T_0=2.25\cdot 10^{-4}$ eV and the modern value of intergalactic magnetic field is $B_0=10^{-10}$ Gs,
we find that the estimate of a magnetic field at the recombination is
$B_r\sim 1.2\cdot 10^{-7}$ Gs $\simeq 2.3 \cdot 10^{-9}$ eV$^2$.
Next we find the relative shift of the recombination temperature due to magnetic field from equation~(\ref{28}),
$\Delta T/T_r\sim 10^{-49}$.
Planck collaboration~\cite{Ade:2015cva} gives the value of a magnetic field at recombimation
$B$(1Mpc) $<0.7\cdot 10^{-9}$ Gs. In such fields the relative shift of the recombination temperature will be
smaller by four orders of magnitude.

Plasma of electrons and protons above phase transition is a
thermodynamic system in the paramagnetic phase. On the other hand,
atomic hydrogen below critical temperature is in a diamagnetic phase.
The free energy is being minimized (the pressure is maximized), and the paramagnetic phase is
thermodynamically preferable. Thus, the temperature of the
transition from plasma phase to the hydrogen phase is decreased as compared
to the case of zero magnetic field.

The consideration of the recombination in the early Universe presented above may be
useful for the consideration of recombination of hydrogen in the supernova explosions,
where strong magnetic fields may exist.

The authors thank Yu.A.Simonov and D.N.Voskresensky for helpful discussions.


\end{document}